\documentstyle[12pt,epsf,graphicx,rotating,epsfig]{article}
 
\newcommand{\be}{\begin{eqnarray}}
\newcommand{\ee}{\end{eqnarray}}

\makeindex

\begin{document}

\title{The Color Glass Condensate and the Glasma: Two Lectures}

\author{Larry McLerran\\
  {\small\it Riken Brookhaven Center and Physics Department}\\
  {\small\it Brookhaven National
  Laboratory, Upton, NY 11973 USA}
}

\maketitle

\begin{abstract}
These two lectures concern the Color Glass Condensate and the Glasma.  These are forms
of matter which might be studied in high energy hadronic collisions.  The Color Glass Condensate
is high energy density gluonic matter.  It constitutes the part of a hadron wavefunction important
for high energy processes.  The Glasma is matter produced from the Color Glass Condensate
in the first instants after a collision of two high energy hadrons.  Both types of matter are associated with
coherent fields.  The Color Glass Condensate is static and related to a hadron  wavefunction,
where the Glasma is transient and evolves quickly after a collision.
I present the properties of such matter, and some aspects of what is known of their properties.
 \end{abstract}

\section{Introduction}

\begin{figure}[ht]
        \includegraphics[width=0.90\textwidth]{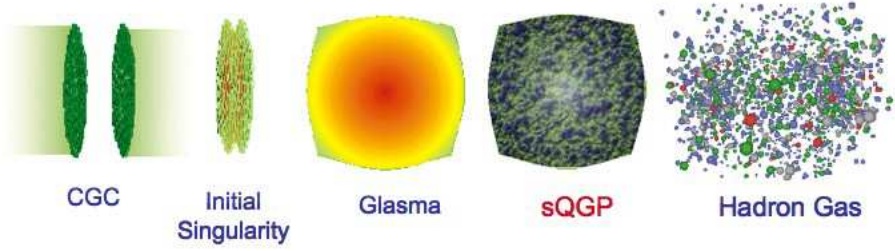}
        \caption{A schematic picture of the evolution of matter produced in the heavy ion collisions. }
\label{lilbang}
\end{figure}

\begin{figure}[ht]
       \begin{center}
        \includegraphics[width=0.90\textwidth]{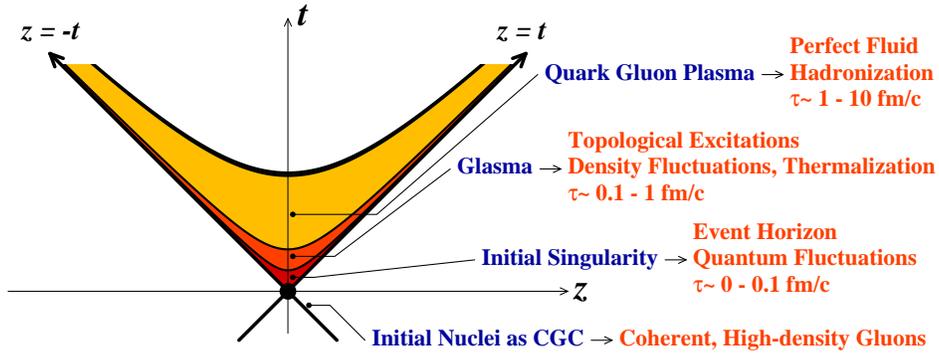}
        \end{center}
        \caption{The various stages in the evolution of the matter produced in high energy hadronic collisions. }
\label{lilbang1}
\end{figure}

There is a generic similarity between the physical conditions in the early stages of heavy ion collisions, and those of the early universe.  Such a collision is illustrated in Fig. \ref{lilbang} . The Color Glass Condensate
describes the initial quantum mechanical wavefunctions which initiate the collision.  This is similar to the initial wavefunction in cosmology.
At the moment of the collision there is the initial singularity, analogous to that of the quantum gravity  and inflationary stages of cosmology, where quantum fluctuations are important. 
These fluctuations  eventually evolve into the large scale fluctuations in the matter
distribution produced in such collisions, analogous to the fluctuations produced in inflationary cosmology which form the seeds for matter fluctuations which eventually develop galaxies and clusters of galaxies. After the singularity there
is a Glasma phase which has topological excitations analogous to those associated with
baryon number violation in electroweak theory. The Quark Gluon Plasma is analogous to both the
electroweak and QCD thermal phases of expansion in cosmology, and the deconfinement transition
is analogous to that which generates masses for electroweak bosons.
In cosmology there are a variety of phase transitions which occur at various times, corresponding to the confinement-deconfinement transition of QCD.     These various stages in the evolution of matter produced in high energy collisions are illustrated in Fig. \ref{lilbang1}.  (The labels of this figure are
slightly modified from the original due to Steffen Bass.)

There is a new paradigm:  Various forms of matter control
the high energy limit of QCD.  These forms of matter have intrinsically interesting properties, as well
as properties which once understood have extensions to other areas of physics such as cosmology.

Several fundamental scientific questions are addressed by a proper understanding of the properties of
these forms of matter.  Among them are
\begin{itemize}
\item{ What is the high energy limit of QCD?}
\item{What are the possible forms of high energy density matter?}
\item{How do quarks and gluons originate in strongly interacting particles?}
\end{itemize}

It is the purpose of these lectures to describe for you the properties of these high energy density forms of matter, how they appear in high energy physics phenomenology, and what is known experimentally about these forms of matter. For reasons of space, these topics are covered only in the most generic terms.


\section{The  Initial Wavefunction}

\begin{figure}[ht]
        \begin{center}
        \includegraphics[width=0.50\textwidth]{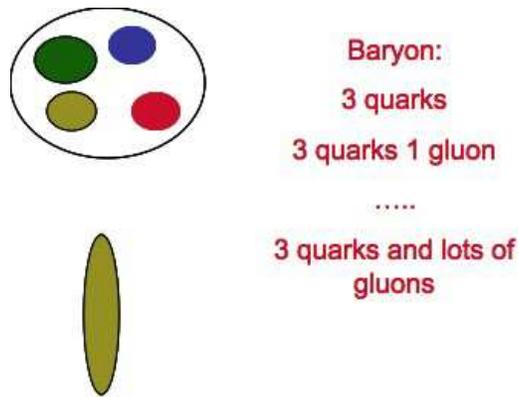}
        \end{center}
        \caption{The leftmost figure on this slide illustrates the wavefunction for a gluon as a function
        of energy.  The rightmost figure illustrates the experimentally measured gluon distribution
        functions as a function of x. }
\label{wfn_glue}
\end{figure}

In the leftmost part of Fig. \ref{wfn_glue},  I show the  various Fock space components for
a baryon wavefunction.  At low energies, the dominant states for physical processes have three quarks and a few gluons.  In high energy collisions, many particles are produced.  These ultimately arise from 
components of the hadron wavefunction which have many gluons in them.  

\begin{figure}[ht]
       \begin{center}
        \includegraphics[width=0.70\textwidth]{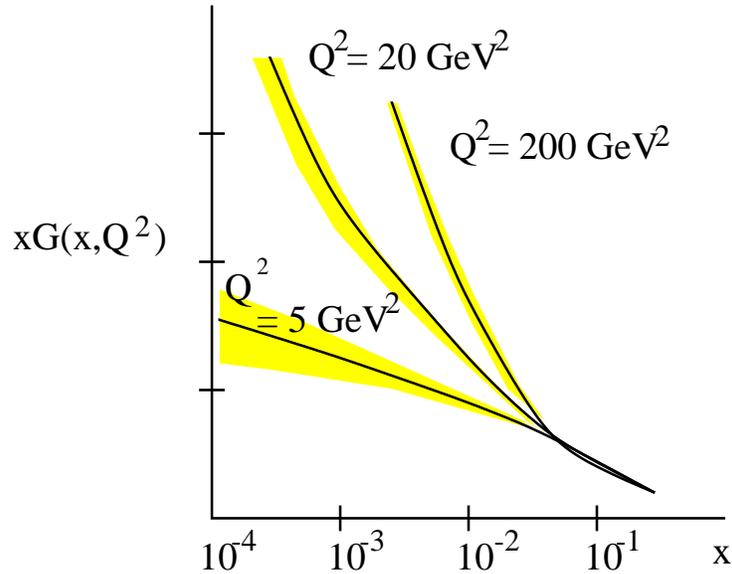}
        \end{center}
        \caption{The distribution of gluons as a function of $x$ as seen in the Hera deep inelastic experiments.}
\label{glue_hera}
\end{figure}

The number of gluons in a hadron wavefunction is usually measured in terms of ratio of the gluon energy to the total energy of a hadron in a frame where the hadron has very high energy,
\begin{equation}
        x = E_{gluon}/E_{hadron}
\end{equation}
Typically, the lowest energy of any gluon is of order $\Lambda_{QCD} \sim 200~MeV$
The minimum value of $x$ is therefore $x_{min} \sim \Lambda/E_{hadron}$  Therefore, the high energy
limit of the description of a hadron is controlled by the low x part of the gluon wavefunction.  The number of gluons have been measured as a function of $x$, and the results of this measurement are shown
in Fig. \ref{glue_hera}.\cite{hera}  The values of $Q^2$ correspond to different resolution scales of
a virtual photon probe.  The main point of the figure is that as $x$ decreases, the number of gluons
increase.  The high energy limit of QCD is therefore described by gluonic states with large numbers of gluons in them.  It is also known from measurements of cross section of strongly interacting particles, that
their size grows very slowly with increasing energy.  Therefore the high energy limit is also the limit
where the gluon density is large.

\begin{figure}[ht]
       \begin{center}
        \includegraphics[width=0.70\textwidth]{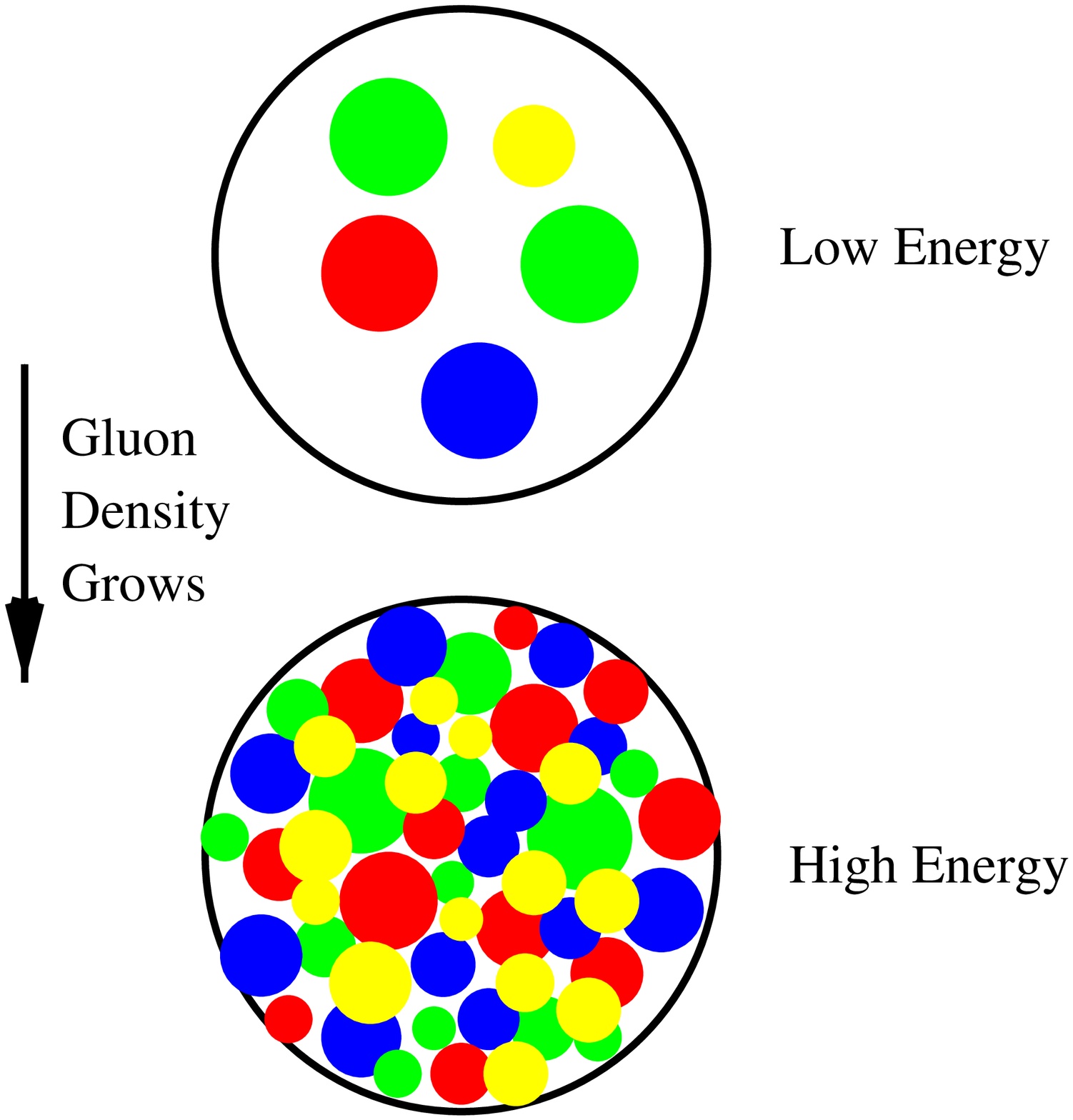}
        \end{center}
        \caption{The density of gluolns in the transverse plane as a function of energy.}
\label{saturation}
\end{figure}

By the uncertainty priincple, these components make
a gluon wall of longitudinal extent of $1/\Lambda_{QCD}$ as shown in Fig. \ref{wfn_glue}  The density of gluons in the
transverse plane grows as the energy increases, as is also shown Fig. \ref{saturation}.


These gluons must become very tightly packed
together.  They form a high density, highly coherent condensate of gluons, the Color Glass Condensate (CGC).\cite{cgc1}-\cite{cgc3}  This coherence arises because the high density implies a high occupancy
of the quantum mechanical gluonic states.  In the high occupancy  limit, the correspondence principle allows the gluons to be described by a classical gluon field.  Classical fields are coherent.  For example, the effects of two positive charges add but the effect of a positive and negative charge cancel.
Because the typical separation of gluons is small, at high enough energy $\alpha_S <<1$,
and the system is weakly coupled. 

Due to the coherence,  the Color Glass Condensate is also strongly interacting.
An example of coherence greatly amplifying a very weak interaction is given by gravity:  The intrinsic
strength of the gravitational force is very small, but due to the coherent superposition of 
forces arising from individual particles, it becomes a large force.
 
There is a typical momentum scale $Q_{sat}$, and gluons with momentum less
than this scale have maximal phase space density,\cite{glr}-\cite{bfkl}
\begin{equation}
	{{dN} \over {d^2p_Td^2r_Tdy}} \sim {1 \over \alpha_S}
\end{equation}
As one goes to higher energies, the saturation momentum increases, as more gluons are added to
the CGC.  This is because once the maximum phase space density is achieved, then repulsive
gluonic interactions forbid greater occupation.  Vacant states of higher momentum must be filled.

We now have all the information needed to understand the word Color Glass Condensate.  Color comes 
from the color of the gluons which make the matter.  Condensate arises from the high occupation
number of the gluons.  The word glass is because the gluons at small x are produced by gluons at
larger versus of x.  In the reference frame where the hadron is fast moving, the high x gluons are
also fast moving.  The natural time evolution is Lorentz dilated.  This dilation is passed on down to 
the scale of the low x gluons, which are then evolving on time scales very large compared to their
natural time scales.  This is the property of a glass.  Ordinary glass is a liquid on very long time scales,
but acts like a solid on shorter time scales.  

Before proceeding further, it is useful to review the properties of some kinematic variables.  These are commonly used in the description of high energy collisions.  We define light cone coordinates
as
\begin{equation}
       x^\pm = (t \pm z)/\sqrt{2}
\end{equation}
and light cone momenta as
\begin{equation}
       p^\pm = (p^0 \pm p^z)/\sqrt{2}
\end{equation}    
The scalar product in these coordinates becomes
\begin{equation}
          x \cdot p = x_T \cdot p_T  - x^+p^- - x^- p^+
\end{equation}
The uncertainty principle for longitudinal light cone coordinates is
\begin{equation}
         x^\pm p^\mp \ge 1
\end{equation}   
It is also useful to define a proper time and a space-time rapidity, useful for $t \ge z$ as
\begin{equation}
  \tau = \sqrt{t^2 - z^2} 
\end{equation}
and
\begin{equation}
     \eta = {1 \over 2} ln \left( {x^+ \over x^-} \right)  
\end{equation}                              
The proper time is invariant under Lorentz boosts along the z axis, and the rapidity transforms by the shift of a constant.  Note that by the uncertainty principle, that a particle will have $M_T = \sqrt{p_T^2 +M^2}
\sim 1/\tau$, and that up to constants of order one,
\begin{equation}
     y = {1 \over 2} ln(p^+/p^-) \sim ln(p^+/M_T) \sim ln(x^+/\tau) \sim \eta
\end{equation}
Thus, all the possible rapidities in momentum space and in coordinate space are of the same order of magnitude.

\begin{figure}[ht]
       \begin{center}
        \includegraphics[width=0.40\textwidth]{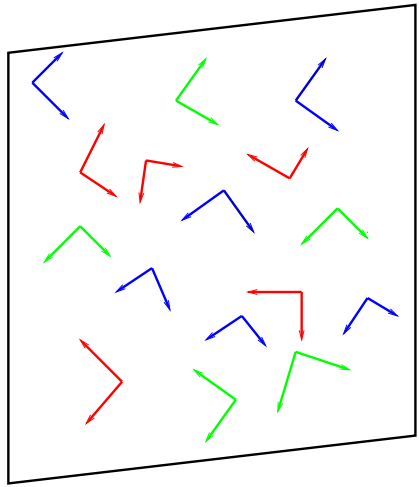}
        \end{center}
        \caption{The color electric and magnetic fields on a sheet of colored glass.}
\label{glasssheet}
\end{figure}

To understand what a sheet of Colored Glass looks like, we can use light cone coordinates.  
Consider $F^{i\pm} \sim E \pm B$, $F^{+-}$, and $F^{ij}$.  The last two terms are of order one.
The terms $F^{i\pm}$ arise from $\mp$ derivatives on the vector potential.  In the  frame where typical
particle momenta are large, $p^+$ is big but $p^-$ is small.  Therefore we expect that 
$F^{i+}$ is big but $F^{i-}$ is small.  This requires that to a first approximation only the transverse
$E$ and $B$ fields are large and that 
\begin{equation}
       E \perp B \perp \hat{z}
\end{equation}        
These are the non-Abelian generalization of the Lienard-Wiiechart potentials of electrodynamics.
The essential complication is that the fields carry a color index which is random on the sheet 
of colored glass.  The density of such fields is determined by saturating the phase space to be of 
order $1/\alpha_S$  This corresponds to a physical number density of gluons of
\begin{equation}
           {1 \over {\pi R^2}} {{dN} \over {dy}} \sim {1 \over \alpha_S} Q_S^2  
\end{equation}           
 In Fig.\ref{glasssheet},  a sheet of colored glass is shown.

\begin{figure}[ht]
       \begin{center}
        \includegraphics[width=0.50\textwidth]{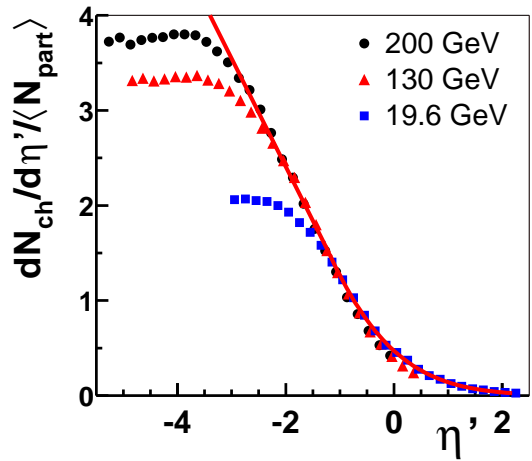}
        \end{center}
        \caption{Limiting fragmentation as illustrated in the Phobos experimental data on gold-gold collisions at RHIC.}
\label{limfrag}
\end{figure}

Limiting fragmentation is a well established phenomenon in high energy scattering.  It is illustrated with
the Phobos experimental data shown in Fig. \ref{limfrag}.\cite{phoboslfrag}-\cite{rajulfrag}  
The total multiplicity of produced particles as a function of rapidity is plotted in this figure.  The data has
been shifted so that the rapidity of one of the nuclei is at zero.  This means rapidities are measured relative to that of one of the nuclei.  For several units of rapidity from that of the nuclei, the multiplicity 
distributions are identical.  When one gets to the central region, corresponding to the smallest x degrees
of freedom accessible in the center of mass frame for the nuclear wavefunction, the scaling disappears.
This is at a smaller value of x as the nuclear energy increases.

The data has a simple interpretation in terms of a renormalization group.  In the figure, imagine we treat
the high x particles as sources.  The fields produced by the sources are those degrees of
freedom in the central region.  Now as we go from lower to higher energy, there are more degrees
of freedom associated with the sources, since the produced fields are localized in a few units 
of rapidity.  Clearly as we go from one energy to another, we must somehow transform what 
were fields into sources.  This happens naturally in a renormalization group formalism.  One simply integrates out dynamical degrees of freedom and converts them to sources as one increases the energy.
This means that the parameters of the theory which describe the system of sources and fields change
as one changes the separation scale between what we call large x and small x degrees of freedom.

Such a theory can be mathematically formulated in terms of a path integral.  Let us treat the large x sources by a current along the light cone,
\begin{equation}
J^\mu = \delta^{\mu+} \rho(x_T,\eta)
\end{equation}
where $\eta$ is the space time rapidity
\begin{equation}
         \eta = ln(x_0^-/x^-)
\end{equation}
The source is assumed to be independent of $x^+$.  This is because of the Lorentz time dilation
of the sources, and reflects the glassy nature of the theory.
The theory is defined by the path integral
\begin{equation}
  Z = \int_\Lambda ~[dA][d\rho]~ exp\left\{ iS[A,\rho] - F[\rho] \right\}
\end{equation}
In this path integral, the action for the gluon fields in the presence of the source $\rho$ is $S[A,\rho]$.
The function $F[\rho]$ is determined by the renormalization group equations.  The function $F$ is real, and this is a consequence of the glassy nature of the system.  The parameter $\Lambda$ is a separation
scale in   longitudinal momenta.  Fields with momenta smaller than $\Lambda$ are included in  the fields $A$, and those with momenta larger than $\Lambda$ are in the sources $\rho$.  

In leading order, the path integral is approximated by doing a stationary phase approximation where the
field $A$ is replaced by its classical value for the equations of motion.  If we do small fluctuations,
these corrections are of the order $\alpha_S ln(p^+/\Lambda)$, where $p^+$ is a typical momentum scale where we measure the classical field.   If $p^+ \sim \Lambda$, then because $\alpha_S << 1$,
these corrections are small, and the classical approximation is valid.  If however, $p^+ >> \Lambda$, then the classical approximation may break down.  In this case, one has to integrate out the quantum
fluctuations successively, to evolve the theory down to a new scale $\Lambda^\prime$, where
$\Lambda^\prime$ is closer to $p^+$.  The integrations and change of scale for the effective theory define the renormalization group.  Remarkably, one can get an explicit form for these renormalization group equations, and show that the effective action is precisely described by the above path integral.      

The renormalization group equations are for the function
\begin{equation}
        Z_0 = e^{-F[\rho]}
\end{equation}
are Hamiltonian equations
\begin{equation}
  {d \over {dy}} Z_0 = - H[d/d\rho, \rho] Z_0
\end{equation}        
Here $y = ln(\Lambda^\prime/\Lambda)$
For strong to intermediate strength fields, $H$ is second order in derivatives.  

It turns out that the Hamiltonian is purely a non-linear second derivative theory.  There is precisely zero
potential.  This means that the solution for $Z_0$ corresponds to diffusion.  The wavefunction spreads as
$y \rightarrow \infty$.  It is this spreading which is the manifestation of the small $x$ problem:\linebreak
{\bf The saturation momentum grows forever as $y \rightarrow \infty$ }

Oftentimes, the source function $F$ is taken to be
\begin{equation}
         F = { 1 \over {2Q_{sat}^2}}~\int~d\eta d^2x_T ~\rho^2(\eta, x_T)
\end{equation}
This model has many of the features of the full solution of the renormalization group equations for $F$.

The solution to the CGC renormalization group equations are  universal, and is a consequence of
the diffusive nature of the evolution.  This means that
as $y \rightarrow \infty$, $F$ tends to a universal function, implying that the properties
of the CGC at arbitrarily high energy are independent of what hadron generated it.  The CGC
is a universal form of matter.

The saturation momentum scale that appears in  $F$, shown explicitly for the Gaussian model,  generates an infrared cutoff scale.  This saturation momentum at high energy can be $Q_S >> \Lambda_{QCD}$
This means that quantities such as the total multiplicity of produced particle become computable
using weak coupling methods.  Typically momentum distributions computed in perturbation theory have infrared divergences at small $p_T$.  However, these become cutoff at the saturation momentum scale, and become perturbatively computable.  For example, the total multiplicity of gluons produced in a collision follows from dimensional reasoning:\cite{nardi}
\begin{equation}
        { 1 \over {\pi R^2}} {{dN} \over {dy}} = {1 \over \alpha_S} Q_{sat}^2
\end{equation}
The factor of $1/\alpha_S$ arises because the multiplicity may be determined by solving a classical     
 problem, or alternatively, because the phase space density of gluons in the initial state has
 the same factor.  We will describe the classical problem in the next lecture about the Glasma.   
 
 The $x$ dependence of the saturation momentum itself can be computed using the above
 renormalization group methods.\cite{mt}  One finds that the saturation momentum grows as
 \begin{equation}
          Q_S^2(y) = Q_0^2 e^{\kappa y}
 \end{equation}
 where $y = ln(1/x)$.  This results is to lowest order in weak coupling.  Running coupling constant
 corrections change this somewhat, but for intermediate values of $x$, the x, dependence is not
 greatly modified.
 
 The Froissart bound on the total cross section also arises very simple.\cite{wiedeman}-\cite{ikeda}  Imagine that
 we have an impact parameter profile of the matter in a hadron.  On general grounds, this should fall
 off exponentially like $e^{-2m_\pi b}$ since the lowest mass iso-singlet channel is due to two pion exchange.  On the other hand, the number of gluons rises like $e^{\kappa y}$.  Therefore a probe
 sees a matter distribution 
 \begin{equation}
 {{dN} \over {d^2bdy}} \sim e^{\kappa y} e^{-2m_{\pi} b}
 \end{equation}         
 A probe of some fixed resolution scale will see the edge of this matter distribution when the
 distribution is some fixed number.  Therefore, the edge is determined by $b \sim \kappa y/2m_\pi$
 Since the maximal $y$ for beam energy $E$ is $y \sim ln(E/E_0)$, we see that the cross section
 saturates the Froissart bound,
 \begin{equation}
   \sigma \sim b^2 \sim ln^2(E/E_0)
 \end{equation}
 
 Consider deep inelastic scattering of a virtual photon from a hadronic target.  The cross section for this
 should scale like $\sigma_{\gamma^* p} \sim F(Q^2/Q_{sat}^2)$  This scaling relation assumes
 the only scale that determines the physics is the saturation momentum.  There is no added $x$ dependence because the local theory which describes the process only has $Q_{sat}$ as a parameter.
\begin{figure}[ht]
       \begin{center}
        \includegraphics[width=0.60\textwidth]{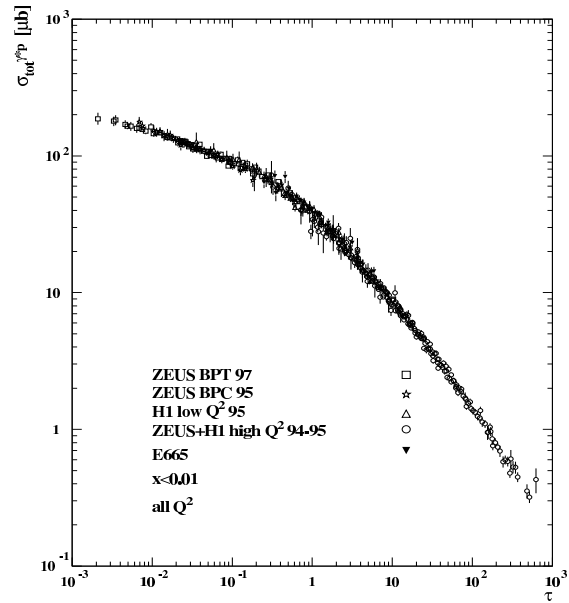}
        \end{center}
        \caption{Geometric scaling in the Hera data..}
\label{geomscaling}
\end{figure}
The geometric scaling of Hera data for this process is shown in Dig. \ref{geomscaling}.\cite{iim}-\cite{gbks}
This scaling works so lang as $x \le 10^{-2}$.  It is also remarkable that the scaling works
to $Q^2 >> Q_{sat}^2$.  The upper limit can be shown to be of order $Q^2 \sim Q_{sat}^4/\Lambda^2_{QCD}$, and introduces the concept of an extended scaling region.

 The theory of the CGC has provided both a rich phenomenology as well as first principles understanding within QCD for the high energy limit.  Some of the successes of this theory include\cite{cgc3}
\begin{itemize}
\item{Scaling properties in electron-hadron scattering.}
\item{Elastic and almost elastic electron-hadron scattering.}
\item{Nuclear size dependence of quark and gluon distributions.}
\item{Distributions of produced particles in hadron and nuclear collisions.}
\item{Scaling properties of hadron-hadron collisions.}
\item{Long range momentum correlations for particles produced in hadron-hadron collisions.}
\item {Intuitive understanding of high energy limit for cross sections.}
\item{The origin of the multiparticle excitations which control high energy scattering (Pomeron, Reggeon,
Odderon).}
\end{itemize}
This is a very active area of research.  It has as one of its intellectual
goals, the unification of the description of all strong interaction processes at high energy
and as such involves different phenomena:  electron-hadron, hadron-hadron, hadron-nucleus and nucleus-nucleus collisions. 

\section{The Initial Singularity and the Glasma}

Before the collision of two hadrons, two sheets of Colored Glass approach one another.\cite{glasma1}-\cite{glasma3}  Because
the phase space density of gluons is large, the gluons can be treated as classical fields.  The fields
are Lorentz boosted Coulomb fields, that is Lienard-Wiechart potentials, which are static in the
transverse plane of the hadrons and have $E \perp B \perp \hat{z}$, where $z$ is the direction of motion.\begin{figure}[ht]
\begin{center}
        \includegraphics[width=0.80\textwidth]{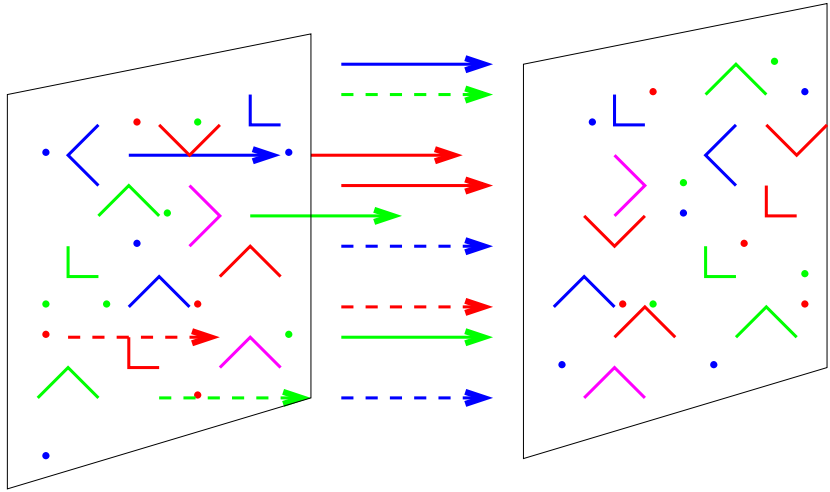}
\end{center}        
        \caption{After the collision of two sheets of colored glass, longitudinal electric and
        magnetic fields associated with the Glasma form. }
\label{glasma}
\end{figure}
When the classical equations for the evolution of these fields is solved, in a very short 
time scale of order
$\Delta t \sim {1 \over Q_s} e^{-\kappa /\alpha_S}$, the fields change from purely transverse to purely
longitudinal.  This is because the collisions generate color electric and magnetic monopole charge
densities of opposite sign on the two sheets.  This description is reminiscent of the  flux tube models of
color electric fields which are used in phenomenological descriptions of low energy strong interactions.
At high energies,  there are both color electric and magnetic field because the electric and magnetic
fields were of equal magnitude in the CGC, and because of the electric-magnetic duality of QCD.

To understand how such fields arise, let us first consider the mathematics of the one nucleus problem.
Assume the sources of the color field are of the form
\begin{equation}
                J^+ = \delta(x^-) \rho(x_T)
\end{equation}
A solution of the classical field equations is for $A^\pm = 0$, and the transverse vector potential
is a pure two dimensional gauge transform of vacuum field, but a different gauge transformation
on opposite sides of the sheet at $x^- = 0$,
\begin{equation}
           A^j =  \theta(-x^-){1 \over i} U_1(x_T )\nabla^j U^\dagger_1(x_T)
           + \theta(x^-){1 \over i} U_2(x_T )\nabla^j U^\dagger_2(x_T)
\end{equation}
This configuration will generate the color electric and magnetic fields described above.

\begin{figure}[ht]
\begin{center}
        \includegraphics[width=0.80\textwidth]{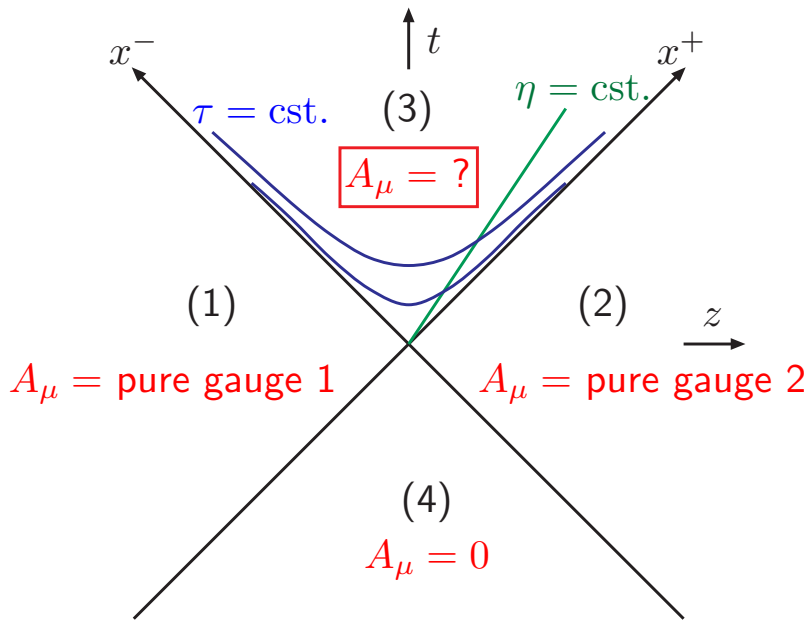}
\end{center}        
        \caption{The fields which describe high energy hadronic collisions. }
\label{fields}
\end{figure}

In the collision problem, the fields in the backward light cone are determined according to
Fig.\ref{fields}.
We have chosen to work in a gauge where as $t \rightarrow -\infty$, the fields vanish.
It is easily checked that when crossing the backwards light cone, one gets the proper discontinuity.
In the forward light cone, one can check that a solution of the equation of motion may be chosen
to be independent of the space-time rapidity $\eta$.  Infinitesmally on the forward light cone,
the vector potential is $A = A_1 + A^2$.  This manes that a color electric and color magnetic charge is induced on the sheets in the forward light cone corresponding to $f_{abc} A^j_{1b} E^j_{2c}$
and $f_{abc} A^j_{1b} B^j_{2c}$ on sheet 2, and with the labels 1 and 2 reversed on sheet 1.

Fields with a non-zero $\vec{E} \cdot \vec{B}$ carry a topological charge.  In QCD, they are associated
with anomalous mass generation and chiral symmetry violation.  In electroweak theory, such
fields may be responsible for generating the baryon asymmetry of the universe.  In QCD, they may
generate the masses of those particles which constitute the visible matter of the universe. Each field
configuration violates CP.  An experimental discovery of the effects of such fields would be of great importance.
Theoretical ideas for experimental signatures are sketchy.\cite{harmen}   These initial longitudinal fields
at the time of  production evolve into transverse fields and eventually form a Quark Gluon Plasma. The matter from production until the formation of the Quark Gluon Plasma is called the Glasma.\cite{glasma1}-\cite{glasma3}

The initial Glasma fields are unstable against forming rapidity dependent fluctuation, and after a time scale of order $1/Q_s$,
instabilities may begin to become of the order of the original classical fields.\cite{mrowcz}-\cite{romatschke} The origin of these Weibel instabilities was originally found
for plasmas close to thermal equilibrium by Mrowczinski.  Quantum fluctuations in the original
wavefunction can grow by these instabilities, and eventually overwhelm the longitudinal electric
and magnetic Glasma fields.  Perhaps these fields form a chaotic or turbulent liquid which might
thermalize and isotropize the system.\cite{muller}

The amplification of quantum fluctuations to macroscopic magnitude is reminiscent of inflation
in the early universe.   These quantum fluctuations expand to size scale larger than the event
horizon during inflation, and are imprinted into the fabric of space-time.  At much later times, the
event horizon size scale becomes of the order of galactic size scales.  Ultimately,  these fluctuations drive gravitationally unstable modes which form galaxies and clusters of galaxies.

These rapidity dependent fluctuations, although they may become large, are in the language of cosmology, sub-horizon, and are modified by the late expansion.  
In addition to these rapidity dependent quantum fluctuations , there may be rapidity independent
fluctuations which are by definition super-horizon.  These are associated with the initial state of the collision and have their origin in the long range Glasma fields.  These would be largely
unaffected by subsequent expansion.  These may be responsible for the
observed forward-backward correlations and perhaps the ridge seen in the two particle
 correlation.\cite{bsrivastava}-\cite{amp}

In heavy ion collisions, analogous fluctuations in the hadronic wavefunction might
appear as momentum dependent fluctuations.  Such
fluctuations might be frozen into the final state distribution of particles.

%%%%%%%%%%%%%%%%%%%%%%%%%%%%%%%%%%%%%%%%%%%%%%%%
%% BACKMATTER
%%%%%%%%%%%%%%%%%%%%%%%%%%%%%%%%%%%%%%%%%%%%%%%%

\section{Acknowledgments}
I gratefully acknowledge my colleagues Edmond Iancu, Dima Kharzeev, Genya Levin, Al Mueller, and Raju Venugopalan for their insights in formulating the material presented here.  

This manuscript has been authorized under Contract No. DE-AC02-98CH0886 with the US Department of Energy.

\end{document}